\documentclass[11pt,a4paper]{article}
\pdfoutput=1
\usepackage{jcappub}
\usepackage{bm}
\usepackage{amsmath}
\usepackage{amssymb}
\usepackage{graphicx}
\usepackage{float}

\begin{document}
\title{Constraint on Brans-Dicke theory from intermediate/extreme mass ratio inspirals}

\author[a]{Tong Jiang,}
\author[a]{Ning Dai,}
\author[a,1]{Yungui Gong\note{Corresponding author.},}
\author[b]{Dicong Liang,}
\author[a]{Chao Zhang}

\affiliation[a]{School of Physics, Huazhong University of Science and Technology, 1037 LuoYu Rd, 
Wuhan, Hubei 430074, China}

\affiliation[b]{Kavli Institute for Astronomy and Astrophysics, Peking University, 5 Yiheyuan Rd, Beijing
100871, China}

\emailAdd{jiangtong@hust.edu.cn}
\emailAdd{daining@hust.edu.cn}
\emailAdd{yggong@hust.edu.cn}
\emailAdd{dcliang@pku.edu.cn}
\emailAdd{chao\_zhang@hust.edu.cn}

\abstract{Intermediate/Extreme mass ratio inspiral (I/EMRI) system provides a good tool
to test the nature of gravity in strong field.
Based on the method of osculating orbits,
we compute the orbital evolutions of I/EMRIs on quasi-elliptic orbits in
both Einstein's general relativity and Brans-Dicke theory.
The extra monopolar and dipolar channels in Brans-Dicke theory accelerate the orbital decay,
so it is important to consider the effects of monopolar and dipolar emissions on the waveform.
With the help of accurate orbital motion,
we generate waveform templates which include both monopolar and dipolar contributions for I/EMRIs on eccentric orbits in Brans-Dicke theory.
With a two-year observation of gravitational waves emitted from I/EMRIs by LISA,
we get the most stringent constraint on the Brans-Dicke coupling parameter $\omega_0>10^6$.
}

\arxivnumber{2107.02700}

\maketitle

\section{Introduction}
    
Since the first direct detection of gravitational wave (GW) event GW150914 \cite{Abbott:2016blz,TheLIGOScientific:2016agk} by the Laser Interferometer Gravitational-Wave Observatory (LIGO) Scientific Collaboration \cite{Harry:2010zz,TheLIGOScientific:2014jea} and Virgo Collaboration \cite{TheVirgo:2014hva},
there have been tens of confirmed GW detections \cite{LIGOScientific:2018mvr,LIGOScientific:2020ibl,LIGOScientific:2021usb,LIGOScientific:2021djp}.
It is well known that in Einstein's general relativity (GR),
GWs propagating with the speed of light have only two tensor polarizations.
However, six possible polarization states are allowed  in general metric theory of gravity \cite{Eardley:1973br,Eardley:1974nw} and
the number of polarization states depends on the particular theory of gravity \cite{Wagoner:1970vr,Liang:2017ahj,Hou:2017bqj,Gong:2018vbo,Gong:2018cgj}.
Therefore, the detected GWs are useful to  understand the nature of gravity and test GR in strong field and nonlinear regions \cite{TheLIGOScientific:2016src,Abbott:2018lct,LIGOScientific:2019fpa,Abbott:2020jks}.
The observation of GW170817 and its electromagnetic counterpart GRB170817A
constrained the speed of GWs as $-3\times 10^{-15}<c_{gw}/c-1\le 7\times 10^{-16}$ \cite{LIGOScientific:2017zic} and this measurement on the propagation speed of GWs was already used to exclude some alternative theories of gravity \cite{Mirshekari:2011yq,Jimenez:2015bwa,Chesler:2017khz,Baker:2017hug,Creminelli:2017sry,Sakstein:2017xjx,Ezquiaga:2017ekz,Green:2017qcv,Nishizawa:2017nef,Arai:2017hxj,Gong:2017kim,Battye:2018ssx,Casalino:2018tcd,Casalino:2018wnc,Visinelli:2017bny}.
	
Brans-Dicke (BD) theory  of gravity is a simple alternative theory of gravity \cite{Brans:1961sx,Dicke:1961gz}.
In BD theory, the BD scalar field $\varphi$ not only takes the role of $G^{-1}$
but also mediates gravity and excites the scalar breathing mode in GWs.
Cosmological observations on the variation of $G$ can constrain BD theory \cite{Avilez:2013dxa,Alonso:2016suf,Amirhashchi:2019jpf},
but the most stringent constraint on BD theory comes from the Cassini measurement
on the Shapiro time delay in the solar system \cite{Bertotti:2003rm} and the result is $\omega_{\text{BD}} > 40000$ \cite{Bertotti:2003rm,Will:2014kxa}.
For a binary system, the orbital period of the system will decrease due to the loss of energy by the emission of GWs.
In BD theory \cite{eardley1975},
the extra dipolar emission channel of GWs can further decrease the orbital period of a binary system \cite{eardley1975,Will:1977wq},
so the measurement on the secular change in the orbital period of a binary can be used to constrain BD theory \cite{eardley1975,Will:1977wq,Will:1989sk,Damour:1992we,Damour:1998jk,Alsing:2011er,Antoniadis:2013pzd,Zhang:2019hos,Seymour:2019tir}.
By using the measurement of the orbital decay from the pulsar-white dwarf binary PSR J1738+0333,
the BD parameter $\omega_0$ was constrained to be $\omega_0> 25000$ \cite{Freire:2012mg}.

The extra energy loss in BD theory makes both the orbital dynamics and the GW waveform of a compact binary system different from those in GR \cite{Mirshekari:2013vb,Lang:2013fna,Lang:2014osa,Sennett:2016klh,Bernard:2018hta,Bernard:2018ivi,Bernard:2019yfz},
so BD theory can also be probed by the observations of GWs \cite{Will:1994fb,Shibata:1994qd,Saijo:1996iz,Will:1997bb,Brunetti:1998cc,Scharre:2001hn,Will:2004xi,Berti:2004bd,Yagi:2009zm,Yunes:2011aa,Arun:2012hf,Barausse:2016eii,Julie:2017ucp,Gnocchi:2019jzp,Carson:2019fxr,Ma:2019rei,Moore:2020rva,Seymour:2020yle}.
However, the orbital evolution and gravitational radiation from binary black holes (BBHs)
are identical in GR and BD theory,
so GW observations of BBHs are unable to distinguish BD theory from GR \cite{Will:1989sk,Yunes:2011aa,Mirshekari:2013vb}.
By using the simple approximate 1.5 post-Newtonian (PN) waveform template with the dipolar correction in the phase
and the Fisher information matrix (FIM) method,
it was estimated that the observation of a $0.7M_\odot$ neutron star (NS) on a quasicircular inspiralling into a $3M_\odot$ black hole (BH) with a signal-to-noise (SNR) of 10 by LIGO/Virgo detectors could give the constraint  $\omega_{\text{BD}}\gtrsim 2000$ \cite{Will:1994fb}.
However, using the Bayesian inference method,
the GW event GW190426\_152155 of a possible $1.5M_\odot$ NS/$5.7M_\odot$ BH binary only gave the constraint  $\omega_{\text{BD}}\gtrsim 10$  \cite{Niu:2021nic}.

If BHs are more massive, then the frequency of emitted GWs is in the millihertz band
and the GWs should be measured by space-based GW detectors like the Laser Interferometer Space Antenna (LISA) \cite{Danzmann:1997hm,Audley:2017drz}, TianQin \cite{Luo:2015ght} and Taiji \cite{Hu:2017mde,Gong:2021gvw}.
One particular interest in space-based GW detections is the stellar-mass BH or NS
captured into inspiral orbits around massive BHs (MBHs),
the Intermediate/Extreme mass ratio inspiral (I/EMRI).
The mass ratio between the compact stellar object and the MBH is about $1:10^2$-$10^4$ for IMRIs and $\lesssim 1:10^4$ for EMRIs.
LISA might detect IMRIs with an event rate $\sim 3-10$ Gpc$^{-3}$ yr$^{-1}$ \cite{Fragione:2017blf}, or 10 IMRIs consisting of BHs with $10^3\,M_\odot$ and $10\,M_\odot$ at any given time \cite{Miller:2001ez},
or a few I/EMRIs consisting of an intermediate-massive BH and a super-massive BH per year \cite{Miller:2004va}.
In EMRIs, the timescale on the modification of the orbit due to the back-reaction from gravitational radiation is much larger than the orbital period,
so it takes the compact object (CO) the last few years to inspiral deep inside the strong field region of the MBH with a speed of a significant fraction of the speed of light and there are $10^5$-$10^6$ GW cycles in the detector band \cite{Barack:2018yvs}.
The emitted GWs from I/EMRIs encode rich information about the spacetime geometry around the MBH
and they can be used to confirm whether the MBH is a Kerr BH predicated by GR.
Using the pattern-averaged waveforms with five parameters and the FIM method,
it was shown that one-year observations of a $1.4M_\odot$ NS on a quasicircular inspiralling into a $10^3 M_\odot$ ($10^4 M_\odot$) BH with a SNR of 10 by LISA could give the constraint  $\omega_{\text{BD}}\gtrsim 244549$ (29906) \cite{Scharre:2001hn},
or $\omega_{\text{BD}}\gtrsim 203772$ (31062) \cite{Will:2004xi}.
These results also showed that the bound on $\omega_{\text{BD}}$ becomes weaker if the central MBH is more massive.
The constraint is expected to be weaker with more parameters.
With the addition of the spin-orbit coupling \cite{Cutler:1994ys},
the bound is reduced significantly by factors of order $10-20$, for example, $\omega_{\text{BD}}\gtrsim 21257$ (3076) for a $1.4M_\odot$ NS/$10^3 M_\odot$ ($10^4 M_\odot$) BH binary \cite{Berti:2004bd}.
Comparing with the results using pattern-averaged waveforms,
the average of $10^4$ randomly distributed binaries from different directions further reduces the bound by a factor of $1.5-4$ \cite{Berti:2004bd,Yagi:2009zm}.
Although the symmetric mass ratio $\eta$ is degenerate with $\omega_\text{BD}$ at the leading Newtonian order,
higher order PN corrections in the GW phase not only break the degeneracy between $\eta$ and $\omega_\text{BD}$ \cite{Will:1994fb},
but also the degeneracy between $\eta$ and the chirp mass \cite{Cutler:1994ys},
leading to the measurements of the two masses of the binary and the bound on $\omega_\text{BD}$.

Some physical scenarios suggest that there are binaries with a significant eccentricity at merger without being circularized by the emission of GWs \cite{Kozai:1962zz,Heggie:1975tg,Wen:2002km,Miller:2002pg,Samsing:2017oij,Rodriguez:2017pec,DOrazio:2018jnv,Hoang:2019kye},
so it is necessary to consider eccentric binaries to probe astrophysical formation channels of binaries \cite{Mroue:2010re,Nishizawa:2016jji,Nishizawa:2016eza,Breivik:2016ddj,Gondan:2017hbp,Gondan:2018khr,Moore:2019vjj,Wu:2020zwr,Lenon:2020oza,Favata:2021vhw}.
Instead of quasicircular orbits,
using the restricted eccentric waveform \cite{Junker:1992kle,Krolak:1995md} for eccentric binaries,
the bound on $\omega_{\text{BD}}$ is reduced by a factor of $\sim 10$ compared with the quasicircular case.
Including the spin-orbit and spin-spin couplings \cite{Blanchet:1995ez},
the spin precession and small eccentricity and taking the average of $10^4$ randomly distributed binaries,
the constraint from a $1.4M_\odot$ NS/$10^3 M_\odot$ BH binary with a SNR of $\sqrt{200}$ by LISA is $\omega_{\text{BD}}\gtrsim 3523$ \cite{Yagi:2009zm}.
Because of the degeneracy among parameters, the constraint becomes weaker with more parameters.
As discussed in \cite{Ma:2019rei}, the degeneracy between the eccentricity and $\omega_\text{BD}$ may deteriorate the accuracy of $\omega_\text{BD}$ for $e\lesssim 0.1$, but for moderate and large eccentricities, the constraint on $\omega_\text{BD}$ becomes stronger. 
Due to the large eccentricity,
the covariances between non-GR and GR parameters are broken, 
and the constraints on non-GR effects become stronger than quasi-circular constraints for large eccentric I/EMRIs \cite{Moore:2020rva}. 
The above results are based on the FIM method and simple waveform templates.
As evident from the discussion above,
the FIM method gave too optimistic constraint on $\omega_{\text{BD}}$ than the Bayesian inference method
because the FIM method models the likelihood as a covariant Gaussian,
so it has limitation \cite{Vallisneri:2007ev,Shuman:2021ruh}
and henceforth we need to apply other method with more accurate waveform template for better parameter estimation.

Because BD theory modifies gravity in the weak field
and the deviations in the energy flux are largest at small velocities,
the constraints on $\omega_{\text{BD}}$ from EMRIs on quasicircular orbits are worse than those derived from comparable-mass binaries \cite{Yunes:2011aa}.
For EMRIs, half the total energy is radiated earlier than 10 years before the final plunge \cite{Barack:2003fp} and monopolar radiation appears when the eccentricity is nonzero \cite{Will:1989sk,Brunetti:1998cc},
so the early inspiral of EMRIs on quasi-elliptic orbits may place stronger bound on BD theory.
Furthermore, a binary with higher eccentricity emits GWs at a spread of GW frequencies that are peaked at higher harmonics of its orbital frequency than a binary on a circular orbit.
Therefore, it is interesting to use I/EMRIs with the small CO moving slowly in a quasi-elliptic orbit to constrain BD theory.

The long inspiral time of EMRIs makes the generation of accurate template waveforms for matched filtering with the numerical relativity method computationally impossible.
However, the problem can be approached based on the expansion in the mass ratio.
To the lowest order, the small CO can be treated as a point like test particle moving in the geodesics of the central MBH.
To higher orders, the gravitational field, the internal structure of the small CO and the back-reaction of gravitational radiation are treated as perturbations.
The waveforms are calculated by solving the Teukolsky equation \cite{Teukolsky:1973ha} and summing all the multipole modes.
The Teukolsky-based waveforms are computationally expensive.
In order to quickly derive the  equation of motion and the waveform template for EMRIs,
the kludge models including the analytical kludge (AK) model \cite{Barack:2003fp} and the numerical kludge (NK) model \cite{Babak:2006uv} were proposed.
The AK model assumes that the small CO moves on a Keplerian orbit with relativistic corrections such as periapsis precession,
Lense-Thirring precession and inspiral from radiation reaction given by analytic post-Newtonian (PN) evolution equations.
It is extremely quick to calculate,
but it dephases relative to the true waveform within hours.
The augmented AK model improves the accuracy of waveform templates \cite{Chua:2015mua,Chua:2017ujo}.
The NK model combines Kerr geodesic with PN orbital evolution which is caused by radiation reaction of GWs,
and numerically integrates the Kerr geodesic equations along the inspiral trajectory. It is more accurate and computationally expensive than the AK model.

On the other hand, the interaction between the small CO and its own gravitational perturbation can be thought as an effective gravitational self-force (SF) driving the radiative evolution of the geodesic orbit of the central MBH.
The perturbative force includes the radiation reaction of GWs and the gravitational effect caused by the small CO,
corresponding to the dissipative and conservative parts, respectively \cite{Barack:2009ux,Poisson:2011nh}.
Great progresses were made on the calculations of SF and applications of the SF method to I/EMRIs \cite{Mino:1996nk,Quinn:1996am,Barack:1999wf,Lousto:1999za,Burko:2000xx,Detweiler:2000gt,Pfenning:2000zf,Barack:2001gx,Barack:2001bw,Barack:2002ku,Keidl:2006wk,Barack:2007jh,Ottewill:2007mz,Detweiler:2008ft,Barack:2009ey,Barack:2010tm,LeTiec:2011bk,Diener:2011cc,Pound:2012nt,Isoyama:2014mja,Zimmerman:2015hua,vandeMeent:2016hel,Pound:2019lzj,Antonelli:2020aeb,Nagar:2022fep,Albertini:2022rfe,VanDeMeent:2018cgn,Barack:2018yvs}.
To integrate the equations of motion
that govern accelerated motion due to the SF in Schwarzschild spacetime,
the method of osculating orbits was proposed \cite{Pound:2007th}.
The SF method can be applied in strong field regions and can generate waveform
templates for EMRIs with a good accuracy \cite{Pound:2007th,Warburton:2011fk,Osburn:2015duj}.
In this paper, we use the SF method to generate waveforms for eccentric I/EMRIs in the framework of GR and BD theory.
We then discuss the constraint on $\omega_{\text{BD}}$ with LISA.
With two-year observations of I/EMRIs,
we derive the constraint $\omega_0>10^6$,
which is more stringent than current solar system tests \cite{Bertotti:2003rm,Will:2014kxa}.
The constraint obtained in this paper is better and should be more robust than those derived in \cite{Scharre:2001hn,Will:2004xi,Berti:2004bd,Yagi:2009zm} with the analytic 2PN waveforms and the FIM approximation.
In this paper we adopt natural units $c=G=1$.
\begin{figure}[htbp]
    \centering
    \includegraphics[width=0.6\columnwidth]{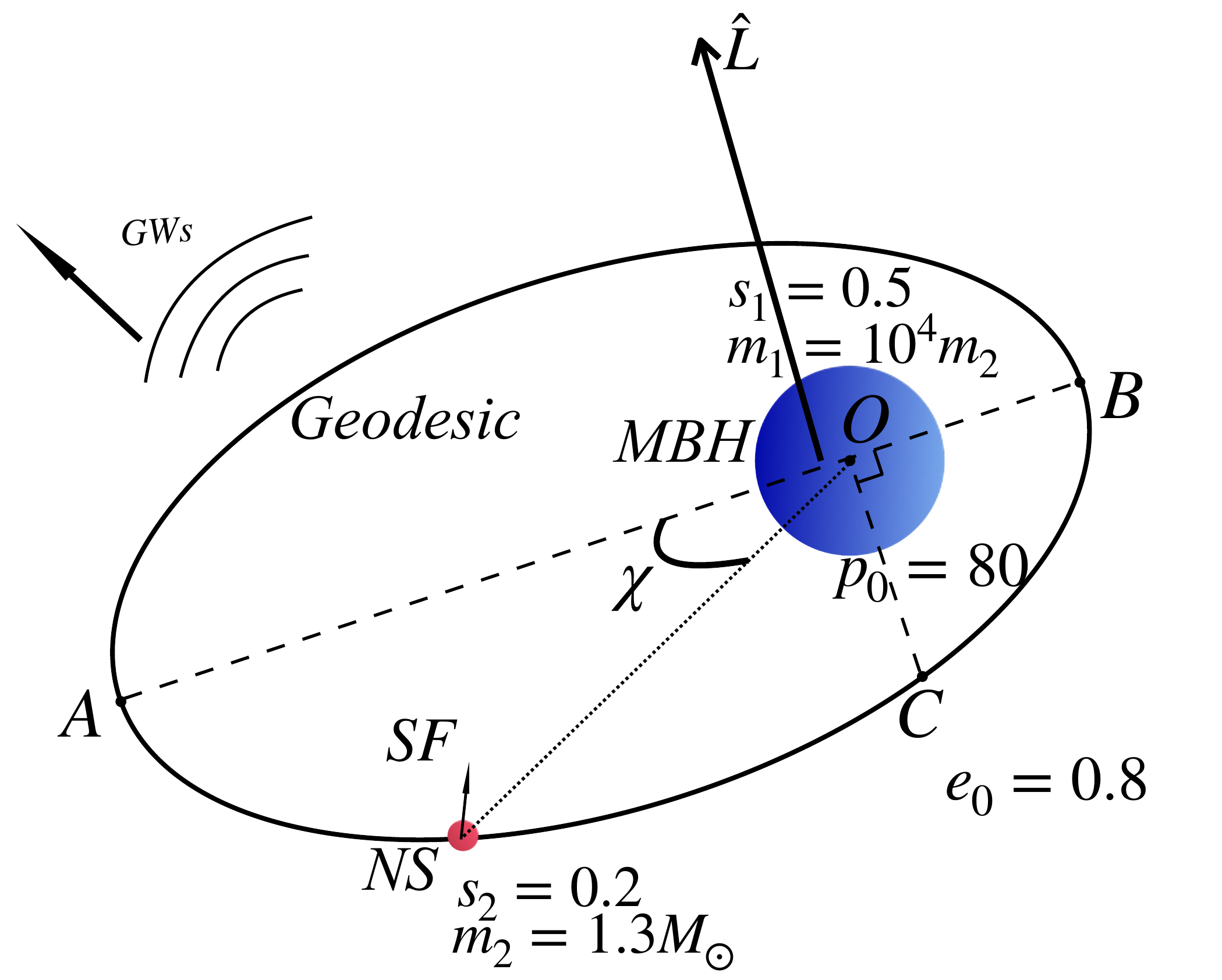}
    \caption{The schematic of the orbit of a binary consisting of a MBH with the mass $m_1=1.3\times 10^4 M_\odot$ and the sensitivity $s_1=0.5$ and a NS with the mass $m_2=1.3 M_\odot$ and the sensitivity $s_2=0.2$ inspiralling around the MBH. The NS follows the geodesic of the central MBH and the SF perturbs the geodesic motion of the NS. The radiation of GWs shrinks the orbit.
    The point A is the apoapsis and the point B is the periapsis.
    $\hat{L}$ is the orbital momentum of the binary, the initial eccentricity $e_0=0.8$, and the initial value of the semilatus rectum (OC) $p_0=80$.
    }
    \label{fig:schematic}
\end{figure}
\section{Overview of the self-force method}
\label{SF}

For I/EMRI systems, we parameterize the bound geodesic by $\chi$ as
\begin{equation}
\label{eq.12}
r(\chi)=\frac{pm_1}{1+e\cos{(\chi-\xi)}},
\end{equation}
where $m_1$ is the mass of the central MBH,
the variation of parameter $\chi$ is $2\pi$ over one radial cycle,
the longitude of pericenter $\xi$ is the value of $\chi$ at periapsis,
$e$ is the orbital eccentricity and $p$ is the semilatus rectum.
Figure \ref{fig:schematic} shows a schematic of I/EMRIs.
At the zeroth order of approximation the small CO moves along the geodesic of the central MBH.
Because of the spherical symmetry of the Schwarzschild BH, the geodesics are in the equatorial plane with $\theta=\pi/2$,
\begin{equation}
\label{eq:9}
    \dot{t}=E/F,
\end{equation}
\begin{equation}
\label{eq.10}
    \dot{r}^2=E^2-U_\text{eff},
\end{equation}
\begin{equation}
\label{eq.11}
    \dot{\phi}=\frac{L}{r^2},
\end{equation}
where the constants $E$ and $L$ correspond to the energy and angular momentum of the system, $F=1-2m_1/r$, the effective potential $U_\text{eff}=F(1+L^2/r^2)$,
and the overdot means a derivative with respect to the proper time $\tau$.
To solve the geodesic eqs. \eqref{eq:9}-\eqref{eq.11}, we use the parameterization 
\eqref{eq.12} and take the parameters $p$, $e$ and $\chi$ as functions of $\tau$.

With the parameterization \eqref{eq.12},
the radial component of the velocity becomes
\begin{equation}
\label{eq.13}
r'(\chi)=\frac{p\,m_1e\sin{(\chi-\xi)}}{\left[1+e\cos{(\chi-\xi)}\right]^2},
\end{equation}
where the prime indicates the derivative with respect to $\chi$.
We can relate the proper time $\tau$ and the parameter $\chi$ using $d\tau/d\chi=r'/\dot{r}$,
so the geodesic eqs. \eqref{eq:9}-\eqref{eq.11} can be parameterized by $\chi$ and they become
\begin{align}
   \label{eq.25}
   \phi'(\chi)=& \sqrt{\frac{p}{p-6-2 e \cos (\chi -\xi)}},\\
   \label{eq.26}
    t'(\chi )=&\frac{m_1 p^2}{[e \cos (\chi -\xi)+1]^2 [p-2-2 e \cos (\chi -\xi )]}\sqrt{\frac{(p-2-2 e) (p-2+2 e)}{p-6-2 e \cos (\chi -\xi)}}.
\end{align}

In terms of the orbital parameters $p$ and $e$,
the energy and the angular momentum of the system are
\begin{equation}
\label{eltoep}
\begin{split}
E^2&=\frac{(p-2-2e)(p-2+2e)}{p(p-3-e^2)},\\
L^2&=\frac{p^2m_1^2}{p-3-e^2}.
\end{split}
\end{equation}

The SF is considered as a perturbation acting on the geodesic of the central MBH.
With the SF, the small CO moves along the worldline $z^\alpha(\lambda)$ parametrized by the affine parameter $\lambda$,
\begin{equation}
\label{eq.1}
    \ddot{z}^\alpha(\lambda)+\Gamma^{\alpha}_{\beta \gamma}\dot{z}^\beta(\lambda)\dot{z}^\gamma(\lambda)=f^{\alpha},
\end{equation}
where the components of the perturbing force $f^\alpha$ are \cite{Pound:2007th}
\begin{equation}
\label{eq.7}
    f^r=\frac{\dot{t}\left[a^r_\epsilon\left(F-r^2\left(\frac{d\phi}{dt}\right)^2\right)+a^\phi_\epsilon r^2\frac{dr}{dt}\frac{d\phi}{dt}\right]} {F^{-1}\left(F^2-\left(\frac{dr}{dt}\right)^2-Fr^2\left(\frac{d\phi}{dt}\right)^2\right)},
\end{equation}
\begin{equation}
\label{eq.8}
    f^\phi=\frac{\dot{t}\left[a^r_\epsilon\frac{dr}{dt}\frac{d\phi}{dt}+a^\phi_\epsilon\left(F^2-\left(\frac{dr}{dt}\right)^2\right)\right]}{F^2-\left(\frac{dr}{dt}\right)^2-Fr^2\left(\frac{d\phi}{dt}\right)^2},
\end{equation}
the subscript $\epsilon$ in the acceleration means that $a^\alpha_\epsilon$ involves only the perturbative terms in $d^2 z^\alpha/dt^2$,
and $a^\alpha_\epsilon$ can be constructed from PN theory.
Take $z^{\alpha}_{G}(I^{A}(\lambda),\lambda)$ as a geodesic with orbital parameters $I^{A}(\lambda)$ and using the osculating condition \cite{Pound:2007th}
\begin{equation}
\label{osccond}
z^\alpha(\lambda)=z^{\alpha}_{G}(I^{A}(\lambda),\lambda),\qquad \frac{dz^{\alpha}(\lambda)}{d\lambda}=\frac{\partial z^{\alpha}_{G}(\lambda)}{\partial \lambda},
\end{equation}
we get the evolution equations for $I^{A}(\lambda)$ as \cite{Pound:2007th}
\begin{equation}
\label{eq.2}
\begin{split}
\frac{\partial z^{\alpha}_{G}}{\partial I^A}\dot{I}^A=0,\\
\frac{\partial \dot{z}^{\alpha}_{G}}{\partial I^A}\dot{I}^A=f^\alpha.
\end{split}
\end{equation}

Explicitly, the osculating orbits \eqref{eq.2} give the evolution equations for the orbital parameters $p$, $e$ and $w$ as \cite{Pound:2007th}
\begin{equation}
\label{eq.4}
\begin{split}
     p'=&\frac{2p^{7/2}m_1^2(p-3-e^2)(p-6-2e\cos Z)^{1/2}
  (p-3-e^2\cos^2Z)}{(p-6+2e)(p-6-2e)(1+e\cos Z)^4} f^{\phi}\\
  &-\frac{2p^3m_1e(p-3-e^2)\sin Z}{(p-6+2e)(p-6-2e)(1+e\cos Z)^2} f^r,
\end{split}
\end{equation}

\begin{equation}
\label{eq.5}
\begin{split}
e'=&\frac{\left\{(p-6-2e^2)
  \left[(p-6-2e\cos Z)e\cos Z+2(p-3)\right]\cos Z+e(p^2-10p+12+4e^2)
  \right\}}{(p-6+2e)(p-6-2e)(p-6-2e\cos Z)^{1/2}
  (1+e\cos Z)^4}\\
  &\times p^{5/2}m_1^2(p-3-e^2) f^{\phi}+\frac{p^2m_1(p-3-e^2)(p-6-2e^2)\sin Z}{(p-6+2e)
  (p-6-2e)(1+e\cos Z)^2} f^r,
\end{split}
\end{equation}

\begin{equation}
\label{eq.6}
\begin{split}
w'=&\frac{p^{5/2}m_1^2(p-3-e^2)\left\lbrace(p-6)
\left[(p-6-2e\cos Z)e\cos Z+2(p-3)\right]-4e^3\cos Z\right\rbrace
  \sin Z}{e(p-6+2e)(p-6-2e)(p-6-2e\cos Z)^{1/2}(1+e\cos Z)^4} f^{\phi}\\
  &-\frac{p^2m_1(p-3-e^2)\left[(p-6)\cos Z+2e\right]}{e(p-6+2e)(p-6-2e)
  (1+e\cos Z)^2} f^r,
\end{split}
\end{equation}
where $Z=\chi-\xi(\chi)$.

Now we apply the above approach to the BD theory with the action \cite{Brans:1961sx,eardley1975}
\begin{equation}
    S=(16\pi)^{-1}\int \left[\varphi R-\varphi^{-1}\omega(\varphi)\varphi^{,\alpha}\varphi_{,\alpha}\right]\sqrt{-g}d^4x
    +S_m(\Psi, g_{\alpha\beta}),
\end{equation}
where $\varphi$ is the BD scalar field, the coupling parameter $\omega$ is a function of $\varphi$ and $S_m$ is the matter action in which the matter field $\Psi$ couples to the metric $g_{\alpha\beta}$ only but the mass of a self-gravitating body depends on the BD scalar field.
We denote the constant $\varphi_0$ as the asymptotic value of $\varphi$ at spatial infinity and $\omega_0=\omega(\varphi_0)$.
The sensitivity of body $A$ is $s_A=d\ln m_A(\varphi)/d\ln\varphi_0$. For BHs, $s=0.5$ \cite{eardley1975}.
In the original BD theory \cite{Brans:1961sx}, $\omega(\phi)=\omega_{\text{BD}}$, the effective Newtonian gravitational coupling constant measured by Cavendish-like experiments is $G=(4+2\omega_{\text{BD}})/[\phi_0(3+2\omega_{\text{BD}})]$.
The stationary, asymptotically flat BHs which are the vacuum solutions in BD theory and GR are the same.

For a compact binary, the relative acceleration between the two bodies up to the 2.5PN in BD theory is \cite{Mirshekari:2013vb}
\begin{equation}
\label{eq.14}
\begin{split}
    \frac{d^2\bm{x}}{dt^2}=&-\frac{\alpha m}{r^2}\bm{n}+\frac{\alpha m}{r^2}
    \left[ \bm{n}(A_{1\text{PN}}+A_{2\text{PN}})+\dot{r}\bm{v}(B_{1\text{PN}}+B_{2\text{PN}}) \right]\\
    &+\frac{8}{5}\eta\frac{(\alpha m)^2}{r^3}[\dot{r}\bm{n}(A_{1.5\text{PN}}+A_{2.5\text{PN}})-\bm{v}(B_{1.5\text{PN}}+B_{2.5\text{PN}})],
\end{split}
\end{equation}
where $\bm{x}=\bm{x}_1-\bm{x}_2$, $r=|\bm{x}|$, $\bm{n}=\bm{x}/r$,
$\bm{v}=\bm{v}_1-\bm{v}_2$, $\dot{r}=dr/dt$,
$m=m_1+m_2$,
\begin{equation}
\label{eq.16}
    \begin{split}
        A_{1\text{PN}}=-(1+3\eta+\bar{\gamma})v^2+\frac{3}{2}\eta\dot{r}^2+2(2+\eta+\bar{\gamma}+\Bar{\beta}_+-\psi\Bar{\beta}_-)\frac{\alpha m}{r},
    \end{split}
\end{equation}
\begin{equation}
\label{eq.17}
    \begin{split}
B_{1\text{PN}}=2(2-\eta+\bar{\gamma}),
    \end{split}
\end{equation}
\begin{equation}
\label{eq.18}
A_{1.5\text{PN}}=\frac{5}{2}\zeta \mathcal{S}^2_{-},
\end{equation}
\begin{equation}
\label{eq.18b}
B_{1.5\text{PN}}=\frac{5}{6}\zeta \mathcal{S}^2_{-},
\end{equation}
\begin{equation}
\label{eq.19}
    \begin{split}
        A_{2\text{PN}}=&-\eta(3-4\eta+\Bar{\gamma})v^4+\frac{1}{2}[\eta(13-4\eta+4\Bar{\gamma})-4(1-4\eta)\Bar{\beta}_++4\psi(1-3\eta)\Bar{\beta}_-]v^2\frac{\alpha m}{r}\\
        &-\frac{15}{8}\eta(1-3\eta)\dot{r}^4+\frac{3}{2}\eta(3-4\eta+\Bar{\gamma})v^2\dot{r}^2+\bigg[2+25\eta+2\eta^2+2(1+9\eta)\Bar{\gamma}+\frac{1}{2}\Bar{\gamma}^2\\
        &-4\eta(3\Bar{\beta}_+-\psi\Bar{\beta}_-)+2\Bar{\delta}_+
        +2\psi\bar{\delta}_-\bigg]\frac{\alpha m}{r}\dot{r}^2
        -\left[9+\frac{87}{4}\eta+(9+8\eta)\Bar{\gamma}+\frac{1}{4}(9-2\eta)\Bar{\gamma}^2\right.\\
        &+(8+15\eta+4\Bar{\gamma})\Bar{\beta}_+-\psi(8+7\eta+4\Bar{\gamma})\Bar{\beta}_-+(1-2\eta)(\Bar{\delta}_+-2\Bar{\chi}_+)+\psi(\Bar{\delta}_-+2\Bar{\chi}_-)\\
        &\left.-24\eta\frac{\Bar{\beta}_1\Bar{\beta}_2}{\Bar{\gamma}}\bigg]\bigg(\frac{\alpha m}{r}\right)^2,
    \end{split}
\end{equation}
\begin{equation}
\label{eq.20}
    \begin{split}
B_{2\text{PN}}=&\frac{1}{2}\eta(15+4\eta+8\Bar{\gamma})v^2-\frac{3}{2}\eta(3+2\eta+2\Bar{\gamma})\dot{r}^2\\
&-\frac{1}{2}[4+41\eta+8\eta^2+4(1+7\eta)\Bar{\gamma}+\Bar{\gamma}^2-8\eta(2\Bar{\beta}_+-\psi\Bar{\beta}_-)+4\Bar{\delta}_++4\psi\Bar{\delta}_-]\frac{\alpha m}{r},
    \end{split}
\end{equation}
\begin{eqnarray}
A_{2.5PN} &=& a_1 v^2 + a_2 \frac{\alpha m}{r} + a_3 \dot{r}^2 \,,
\nonumber \\
B_{2.5PN} &=& b_1 v^2 + b_2 \frac{\alpha m}{r} + b_3 \dot{r}^2 \,,
\label{25pnAB}
\end{eqnarray}
\begin{subequations}
\begin{eqnarray}
a_1 &=& 3 - \frac{5}{2} \bar{\gamma}  + \frac{15}{2} \bar{\beta}_+   +\frac{5}{8} \zeta {\cal S}_{-}^2 (9 + 4\bar{\gamma} -2\eta)
+ \frac{15}{8} \zeta \psi  {\cal S}_{-} {\cal S}_{+} \,,
\\
a_2 &=& \frac{17}{3}  + \frac{35}{6} \bar{\gamma} - \frac{95}{6} \bar{\beta}_+
- \frac{5}{24}\zeta {\cal S}_{-}^2  \left [ 135 + 56\bar{\gamma} + 8\eta + 32\bar{\beta}_+  \right ]
 +30 \zeta {\cal S}_{-} \left ( \frac{{\cal S}_{-} \bar{\beta}_+ + {\cal S}_{+} \bar{\beta}_{-}}{\bar{\gamma}} \right )
 \nonumber \\
&&
-\frac{5}{8} \zeta \psi  {\cal S}_{-} \left ( {\cal S}_{+} - \frac{32}{3} {\cal S}_{-} \bar{\beta}_{-} +16 \frac{{\cal S}_{+} \bar{\beta}_+ + {\cal S}_{-} \bar{\beta}_{-}}{\bar{\gamma}} \right )  -40 \zeta \left (\frac{{\cal S}_{+} \bar{\beta}_+ + {\cal S}_{-} \bar{\beta}_{-}}{\bar{\gamma}} \right )^2 \,,
\\
a_3 &=& \frac{25}{8}  \left [ 2\bar{\gamma} - \zeta  {\cal S}_{-}^2 (1-2\eta)
  - 4\bar{\beta}_+ - \zeta \psi {\cal S}_{-} {\cal S}_{+}\right ]  \,,
\\
b_1 &=& 1 - \frac{5}{6}\bar{\gamma} +\frac{5}{2} \bar{\beta}_+
-\frac{5}{24} \zeta {\cal S}_{-}^2 (7 + 4\bar{\gamma} -2\eta)
+ \frac{5}{8} \zeta \psi  {\cal S}_{-} {\cal S}_{+} \,,
\\
b_2 &=& 3 + \frac{5}{2} \bar{\gamma} -\frac{5}{2} \bar{\beta}_+
 -\frac{5}{24} \zeta   {\cal S}_{-}^2  \left [ 23 + 8\bar{\gamma} - 8\eta + 8\bar{\beta}_+  \right ]\\
 &&
 +\frac{10}{3} \zeta {\cal S}_{-}  \left ( \frac{{\cal S}_{-} \bar{\beta}_+ + {\cal S}_{+} \bar{\beta}_{-}}{\bar{\gamma}} \right )
  \nonumber \\
 &&
 -\frac{5}{8} \zeta \psi  {\cal S}_{-} \left ( {\cal S}_{+} - \frac{8}{3} {\cal S}_{-} \bar{\beta}_{-} + \frac{16}{3} \frac{{\cal S}_{+} \bar{\beta}_+ + {\cal S}_{-} \bar{\beta}_{-}}{\bar{\gamma}} \right ) \,,
 \\
b_3 &=& \frac{5}{8} \left [ 6\bar{\gamma} + \zeta  {\cal S}_{-}^2 (13 + 8\bar{\gamma}+2\eta)
  - 12\bar{\beta}_+  - 3 \zeta \psi {\cal S}_{-} {\cal S}_{+}  \right ]
  \,.
\end{eqnarray}
\label{25PNcoeffs}
\end{subequations}
\begin{equation}
\label{pardef10}
\begin{split}
\psi=\frac{m_1-m_2}{m_1+m_2}=\sqrt{1-4\eta},\\
\mathcal{S_+}=-\alpha^{-1/2}(s_1-s_2), \\
\mathcal{S_-}=-\alpha^{-1/2}(1-s_1-s_2),
\end{split}
\end{equation}
$v=|\bm{v}|$, $\eta=m_1m_2/m^2$, 
$\alpha=1-\zeta+\zeta(1-2s_1)(1-2s_2)$, $\zeta=1/(4+2\omega_0)$, 
$\lambda_1=\lambda_2=0$,
and the other parameters in the above equations are defined in table \ref{table1}.
Here the subscripts ``+" and $``-"$ on various parameters denote the sum and the difference, such as
\begin{equation}
\label{parpmdef}
x_+=\frac{1}{2}(x_1+x_2),\qquad x_-=\frac{1}{2}(x_1-x_2).
\end{equation}

From now on, the overdot means the derivative with respective to $t$.
Note that higher order PN corrections break the degeneracies among $m_1$, $m_2$ and $\omega_\text{BD}$, 
so the two masses $m_1$, $m_2$ and the BD coupling parameter $\omega_\text{BD}$
can be measured at the 2.5PN.
For the equations of motion through 2PN and 3PN, please refer to \cite{Lang:2013fna,Lang:2014osa,Bernard:2018hta}.
In the GR limit $\omega_0\rightarrow \infty$, eq. \eqref{eq.14} reduces to that of GR.
The first term in eq. \eqref{eq.14} is Newtonian gravity, and the presence of $\alpha$ shows the violation of the strong equivalence principle in BD theory.
If the mass of an self-gravitating body is independent of the BD scalar field,
then the sensitivity $s=0$ and $\alpha=1$, we recover the Newtonian gravity.
Since the sensitivity of BHs is $s=1/2$, i.e., $s_1=1/2$ and $s_1'=s_1''=0$, so for BBHs,
the equation of motion \eqref{eq.14} in BD theory is the same as that in GR through 2.5PN if we
rescale each mass by $\alpha$.
In the extreme mass ratio limit, it was found that there is no dipolar radiation to all orders in PN theory for BBHs \cite{Yunes:2011aa}.

\begin{table}[htpb]
\label{table1}
\begin{tabular}{clcl}
Parameter&Definition&Parameter&Definition\\
\hline
\multicolumn{2}{l}{\bf Scalar-tensor parameters}&\multicolumn{2}{l}{\bf Equation of motion parameters}\\
\multicolumn{2}{l}{$G$\qquad\qquad$\phi_0^{-1} (4+2\omega_0)/(3+2\omega_0)$}&\multicolumn{2}{l}{\bf Newtonian}\\
\multicolumn{2}{l}{$\zeta$\qquad\qquad\qquad$1/(4+2\omega_0)$}&\multicolumn{2}{l}{$\alpha $\qquad$1 - \zeta + \zeta (1-2s_1)(1- 2s_2) $}
\\
\multicolumn{2}{l}{$\lambda_1$\qquad\qquad$(d\omega/d(\varphi/\varphi_0))_0 \zeta^2/(1-\zeta)$}&\multicolumn{2}{l}{\bf post-Newtonian}\\
\multicolumn{2}{l}{$\lambda_2$\qquad\qquad$(d^2\omega/d(\varphi/\varphi_0)^2)_0 \zeta^3/(1-\zeta)$}&\multicolumn{2}{l}{$\bar{\gamma}$ \qquad $-2 \alpha^{-1}\zeta (1-2s_1)(1-2s_2)$}
\\
\multicolumn{2}{l}{\bf Sensitivities}&\multicolumn{2}{l}{$\bar{\beta}_1 $\qquad$\alpha^{-2} \zeta (1-2s_2)^2 \left ( \lambda_1 (1-2s_1) + 2 \zeta s'_1 \right )$}
\\
\multicolumn{2}{l}{$s_A$\qquad\qquad$[d \ln m_A(\varphi)/d \ln \varphi]_0$}&\multicolumn{2}{l}{$\bar{\beta}_2 $\qquad$\alpha^{-2} \zeta (1-2s_1)^2 \left ( \lambda_1 (1-2s_2) + 2 \zeta s'_2 \right )$}
\\
\multicolumn{2}{l}{$s'_A$\qquad\qquad$[d s_A/d \ln \varphi]_0$}&\multicolumn{2}{l}{\bf 2nd post-Newtonian}\\
\multicolumn{2}{l}{$s''_A$\qquad\qquad$[d^2 s_A/d \ln \varphi^2]_0$}&\multicolumn{2}{l}{$\bar{\delta}_1$ \qquad$ \alpha^{-2} \zeta (1-\zeta) (1-2s_1)^2$}
\\
&&\multicolumn{2}{l}{$\bar{\delta}_2$ \qquad$\alpha^{-2} \zeta (1-\zeta) (1-2s_2)^2 $}
\\
\qquad\qquad$\bar{\chi}_1 $ & \multicolumn{3}{l}{\qquad\qquad\qquad$ \alpha^{-3} \zeta (1-2s_2)^3$ $  \left [ (\lambda_2 -4\lambda_1^2 + \zeta \lambda_1 ) (1-2s_1) -6 \zeta \lambda_1 s'_1 + 2 \zeta^2 s''_1\right ]$}
\\
\qquad\qquad$\bar{\chi}_2 $&  \multicolumn{3}{l}{\qquad\qquad\qquad$ \alpha^{-3} \zeta (1-2s_1)^3 $ $ \left [ (\lambda_2 -4\lambda_1^2 + \zeta \lambda_1 ) (1-2s_2) -6 \zeta \lambda_1 s'_2 + 2 \zeta^2 s''_2 \right ]$}
\end{tabular}
\caption{\label{tab:params} Parameters used in the equations of motion. We use similar notation as in \cite{Mirshekari:2013vb}.}
\end{table}

For convenience, we rearrange eq. \eqref{eq.14} as
\begin{equation}
\label{eq.15f}
\frac{d^2\bm{x}}{dt^2}=-\frac{\alpha m}{r^2}(A_{BD}\bm{n}-B_{BD}\bm{v}),
\end{equation}
where
\begin{equation}
\label{eq.15g}
\begin{split}
A_{BD}&=1-A_{\text{1PN}}-A_{\text{2PN}}-\frac{8}{5}\eta\frac{\alpha m}{r}\dot{r}(A_{1.5\text{PN}}+A_{2.5\text{PN}}),\\
B_{BD}&=(B_{1\text{PN}}+B_{2\text{PN}})\dot{r}-\frac{8}{5}\eta\frac{\alpha m}{r}(B_{1.5\text{PN}}+B_{2.5\text{PN}}).
\end{split}
\end{equation}
To separate the perturbation from the Schwarzschild background and derive the acceleration $a^\alpha_\epsilon$, we write
$A_{BD}=A_s+\tilde{A}$ and $B_{BD}=B_s+\tilde{B}$,
where the coefficients $A_s$ and $B_s$ coming from the geodesic equations of the Schwarzschild BH through 2.5PN are
\begin{equation}
\label{eq.21}
    A_s=1-4\frac{\alpha m_1}{r}+v^2+9(\frac{\alpha m_1}{r})^2-2\frac{\alpha m_1}{r}(\frac{dr}{dt})^2,
\end{equation}
\begin{equation}
\label{eq.22}
    B_s=-\frac{dr}{dt}(4-2\frac{\alpha m_1}{r}),
\end{equation}
and $\tilde{A}$ and $\tilde{B}$ denote the contributions from perturbations. Combining eqs. \eqref{eq.15f}-\eqref{eq.22}, we get the perturbed accelerations
\begin{equation}
\label{eq.23}
    a^r_\epsilon=-\frac{m_1}{r^2}(\Tilde{A}+ \Tilde{B}\frac{dr}{dt}),
\end{equation}
\begin{equation}
\label{eq.24}
    a^\phi_\epsilon=-\frac{m_1}{r^2} \Tilde{B}\frac{d\phi}{dt}.
\end{equation}
Substituting eqs. \eqref{eq.23} and \eqref{eq.24} into
eqs. \eqref{eq.7} and \eqref{eq.8},
we derive the perturbing force in BD theory
and then we solve the evolution eqs. \eqref{eq.4}-\eqref{eq.6} numerically for the orbital parameters.

\begin{figure}[htbp]
\centering
    \includegraphics[width=0.9\columnwidth]{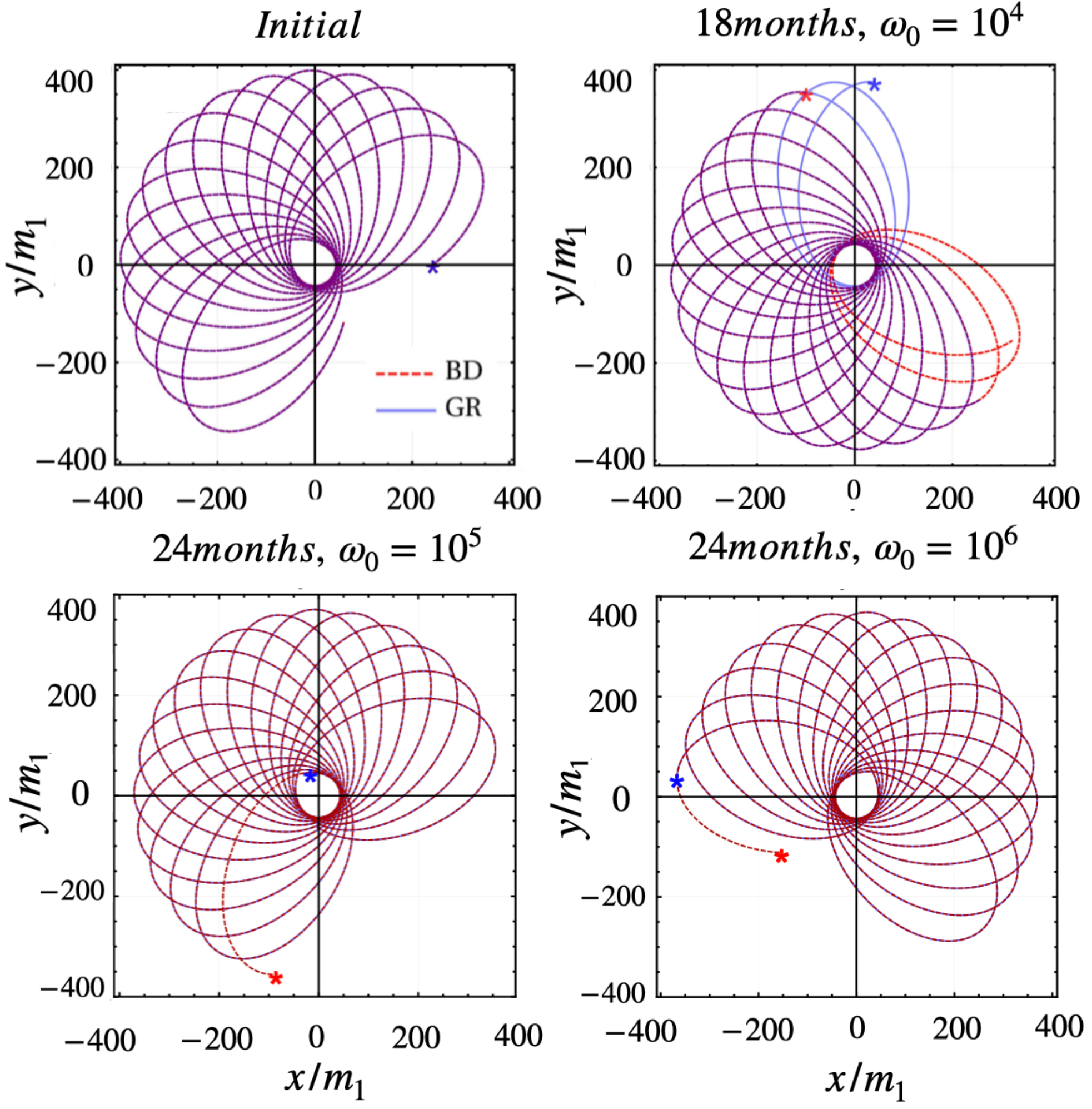}
    \caption{Comparisons of the orbital evolution in GR (blue solid curves) and BD theory (red dashed curves) with different $\omega_0$.
     The upper left panel displays the initial behaviors of the orbits and the other panels show the behaviors of the orbits after 18 and 24 months. ``*" labels the starting position of the compact object. All the plots cover the same range of time (2000 seconds).}
    \label{fig:1}
    \end{figure}

\begin{figure}[htbp]
\centering
    \includegraphics[width=0.6\columnwidth]{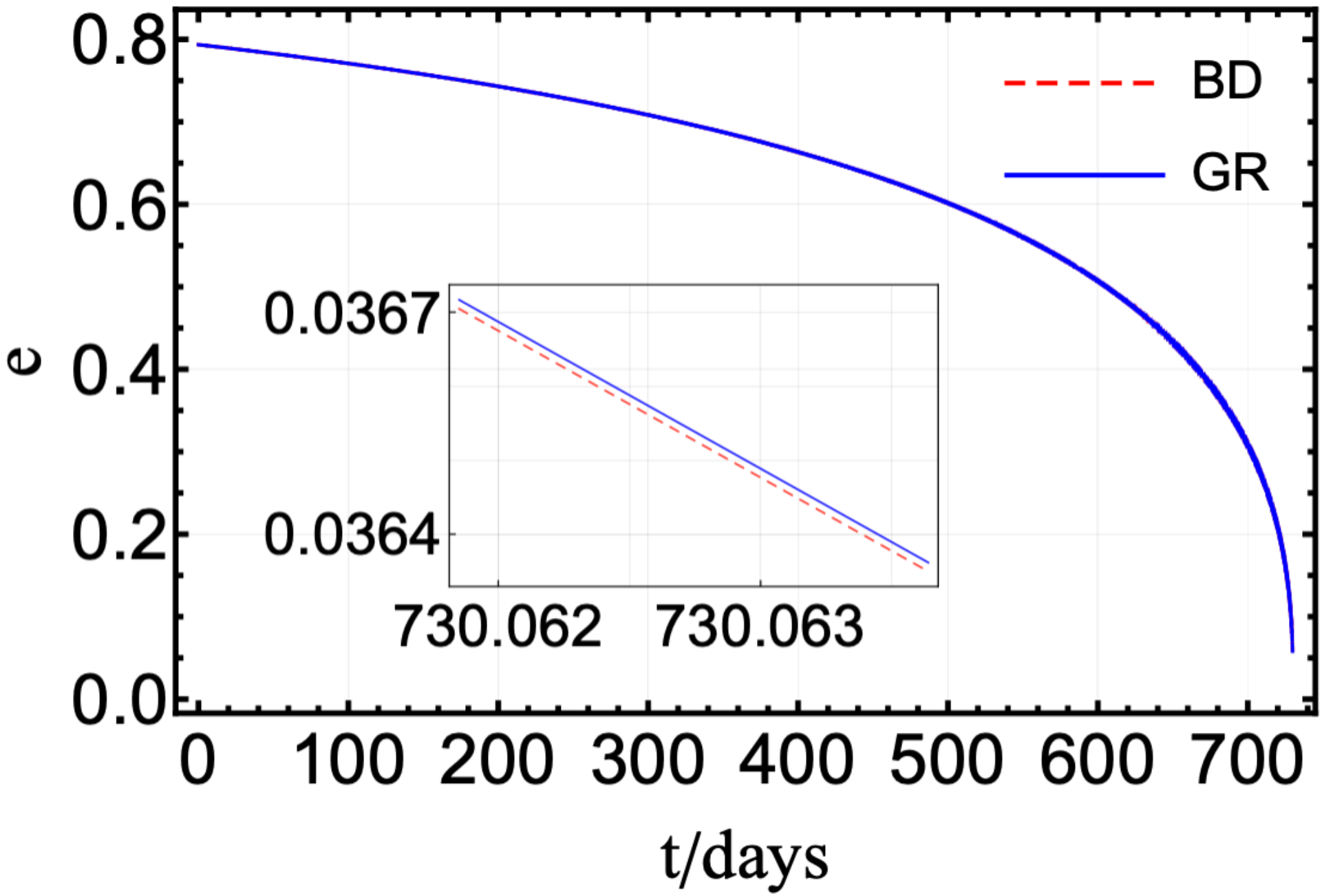}
     \caption{The evolution of the eccentricty in GR (blue solid curves) and BD theory (red dashed curves) with $\omega_0=10^6$.
    The inset shows the evolution of eccentricity in a short time.}
    \label{fig:e}
    \end{figure}

As an example, we consider a binary consisting of a NS and a BH.
We take the mass of the NS as $m_2=1.3M_\odot$, the sensitivity as $s_2=0.2$ \cite{Scharre:2001hn}, 
and the mass ratio between the NS and the BH as $q=m_2/m_1=10^{-4}$.
With these choices of the parameters, we get $s_2'=s_2''=0$, and the PN parameters $\bar{\gamma}=\bar{\beta}_1=\bar{\beta}_2=\bar{\delta}_1=\bar{\chi}_1=\bar{\chi}_2=0$.
The initial conditions are as follows:
at the initial time $t_0=0$, the dimensionless semilatus rectum $p_0=80$,
the eccentricity $e_0=0.8$ \cite{Barack:2003fp,Osburn:2015duj},
the orbital parameter at periapsis $\xi_0=10$,
and the phase at periapsis $\phi_0=0$.
The results for the orbital evolution are shown in figures \ref{fig:1} and \ref{fig:e}.
Figure \ref{fig:1} plots the orbital evolutions and figure \ref{fig:e}
plots the evolution of the eccentricity $e$.
It is shown that
the difference of the evolution of the eccentricity between BD and GR is very small.
We also considered the cases with $q=10^{-3}$ and $q=10^{-6}$.

\begin{figure}[htp]
    \centering
    \includegraphics[width=0.6\columnwidth]{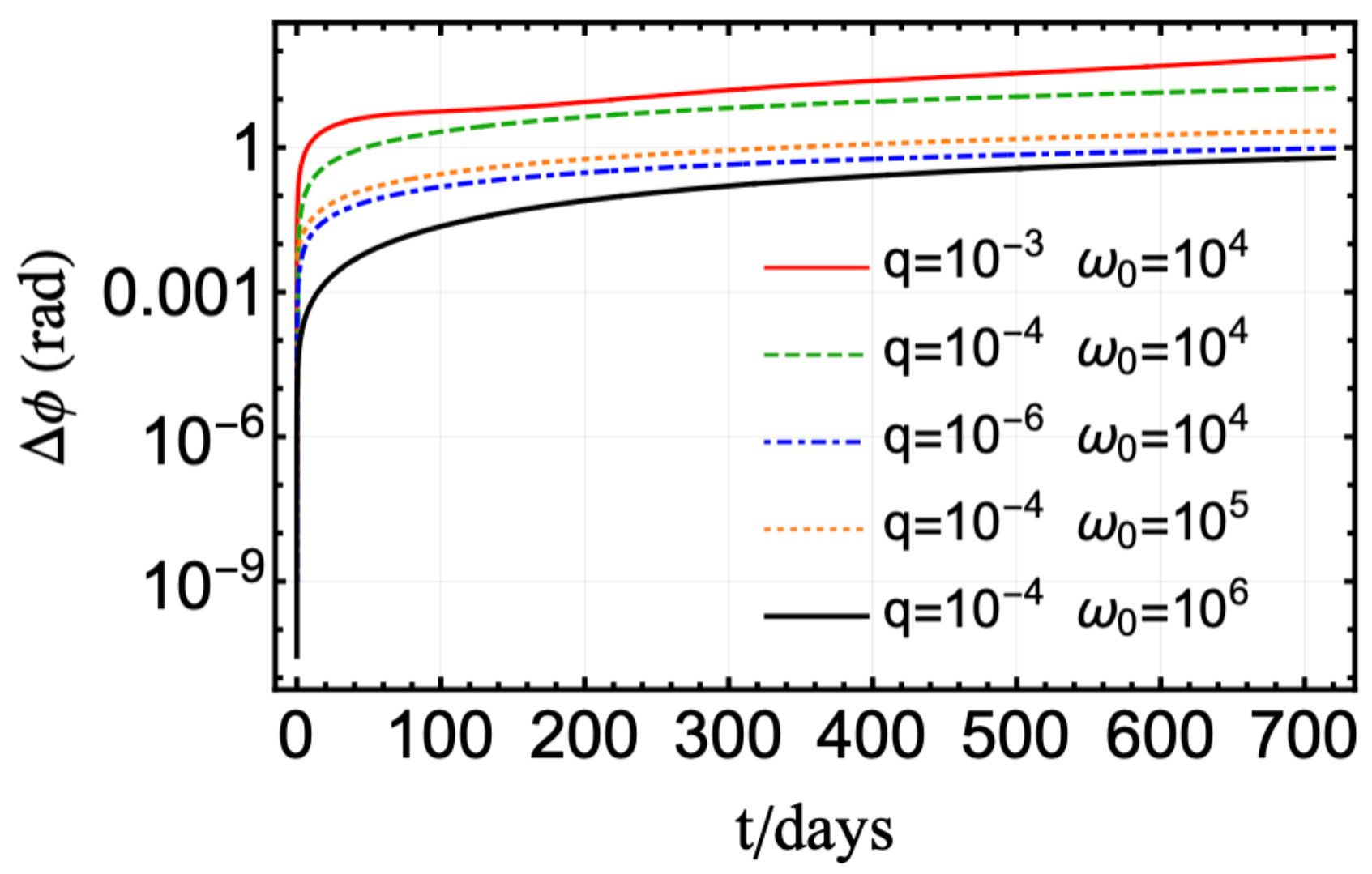}
    \caption{The orbital phase difference $\Delta\phi=\langle\phi_\text{BD}\rangle-\langle\phi_\text{GR}\rangle$  between GR and BD theory with different $\omega_0$ and $q$.}
    \label{dphit}
\end{figure}

From figure \ref{fig:1}, we see that during the initial inspirals,
the orbits in GR and BD theory are almost identical.
The orbits in the BD theory with $\omega_0=(10^4, 10^5, 10^6)$ start to deviate from those in GR after $t=(18, 24, 24)$ months of inspirals, respectively.
The main difference in the orbital evolution can be manifested by
the orbital phase accumulation.
In figure \ref{dphit}, we show the evolutions of the orbital phase difference $\Delta\phi=\langle\phi_\text{BD}\rangle-\langle\phi_\text{GR}\rangle$ between GR and BD theory with different $\omega_0$ and $q$, here $\langle\phi\rangle$ means the phase average over one cycle.
As expected, the phase difference increases as $\omega_0$ becomes smaller
which is consistent with the results of the orbital evolution.
After around one-year inspiral,
the phase difference is 273.4 rad for $\omega_0=2000$ and $q=10^{-3}$;
the phase difference is larger than 40 rad for $\omega_0=2000$ and $q=10^{-4}$;
and the phase difference is 1.62 rad for $\omega_0=50000$ and $q=10^{-4}$.
After around two-year inspiral, the phase difference is 81.93 rad for $\omega_0=10^4$ and $q=10^{-3}$;
the phase difference is larger than 17.24 rad for $\omega_0=10^4$ and $q=10^{-4}$;
the phase difference is 2.24 rad for $\omega_0=10^5$ and $q=10^{-4}$;
and the phase difference is 0.94 rad for $\omega_0=10^6$ and $q=10^{-4}$.
As the mass of the central BH increases,
the accumulated phase difference decreases.
Note that the difference of the number of cycles the NS inspiralled between GR and BD theory is $\Delta N=\Delta\phi/(2\pi)$.
If we can observe the motion of the NS, then we can use $\Delta N$ to distinguish BD theory from GR.
The accumulated phase difference will be manifested in the GW waveform
and make it possible to constrain BD.

\section{Gravitational waveform}
\label{Waveform}
For GR, the waveform up to quadrupole radiation is 
\begin{equation}
\label{hijgr}
       h^{ij}_{\text{GR}}=\frac{4\eta m}{D_L}\left(v^iv^j-\frac{m}{r}n^in^j\right),
\end{equation}
and for BD, the waveform up to quadrupole radiation is \cite{Will:1994fb}
\begin{equation}
\label{hijbd}
    h^{ij}_{BD}=-\frac{2\zeta\mathcal{A}}{D_L}(\delta^{ij}-\Omega^i\Omega^j)
    +\frac{4(1-2\zeta)\eta m}{D_L}(v^iv^j-\frac{m}{r}n^in^j),
\end{equation}
where $D_L$ is the luminosity distance between the source and the detector
and the unit vector $\bm{\Omega}$ is the propagation direction of GWs.
The other parameters are
\begin{equation}
\label{eq.29}
    \mathcal{A}=\mathcal{E}+\dot{\mathcal{E}}^j\Omega_j-\frac{1}{2}\ddot{I}^{jk}\Omega_j\Omega_k,
\end{equation}
\begin{equation}
\label{eq.30}
\mathcal{E}=2(1+2\lambda)\frac{m_1 m_2}{r}+\eta m\left[v^2+(1+4\lambda)\frac{m}{r}\right],
\end{equation}
\begin{equation}
\label{eq.31}
    \mathcal{E}^j=-2(1+2\lambda)\eta m \mathcal{S}r^j-\eta\Delta m\left[v^2+(1+4\lambda)\frac{m}{2r}\right]r^j,
\end{equation}
\begin{equation}
\label{eq.32}
    \ddot{I}^{jk}=2\eta m\left(v^iv^j-\frac{m}{r}n^in^j\right),
\end{equation}
where $\Delta=(m_1-m_2)/(m_1+m_2)$ and
$\mathcal{S}=(m_1-m_2)/(2r)$.
For GWs propagating along the $z$ direction with the unit vector $e_Z$ in the detector-adapted frame,
the polarizations of GWs are expressed as 
\begin{equation}
\label{eq.33}
    h_{+}=\frac{1}{2}h_{ij}(e^i_Xe^j_X-e^i_Ye^j_Y),
\end{equation}
\begin{equation}
\label{eq.34}
    h_{\times}=\frac{1}{2}h_{ij}(e^i_Xe^j_Y+e^j_Xe^i_Y),
\end{equation}
\begin{equation}
\label{eq.35}
    h_{b}=\frac{1}{2}h_{ij}(e^i_Xe^j_X+e^i_Ye^j_Y),
\end{equation}
where the unit vectors $e_X$ and $e_Y$ are perpendicular to  the unit vector $e_Z$ and they along with $e_Z$ form an orthonormal basis.
In the absence of the BD scalar field,  GR is recovered, $h_+$ and $h_{\times}$ are the plus and cross polarizations in the transverse-tracefree (TT) gauge.
In particular, in the heliocentric coordinate system,
\begin{equation}
\begin{split}
    e_X&=[\cos\xi,\sin\xi,0], \\
    e_Y&=[-\cos\iota\sin\xi,\cos\iota\cos\xi,\sin\iota],
\end{split}
\end{equation}
where the inclination angle $\iota$ measures the angle between the propagation direction of GWs and the normal vector of the orbital plane,
the longitude of pericenter $\xi$ is the angle between the pericenter and the line of nodes as measured in the orbital plane.

\begin{figure*}[htp!]
    \centering
     \includegraphics[width=0.95\linewidth]{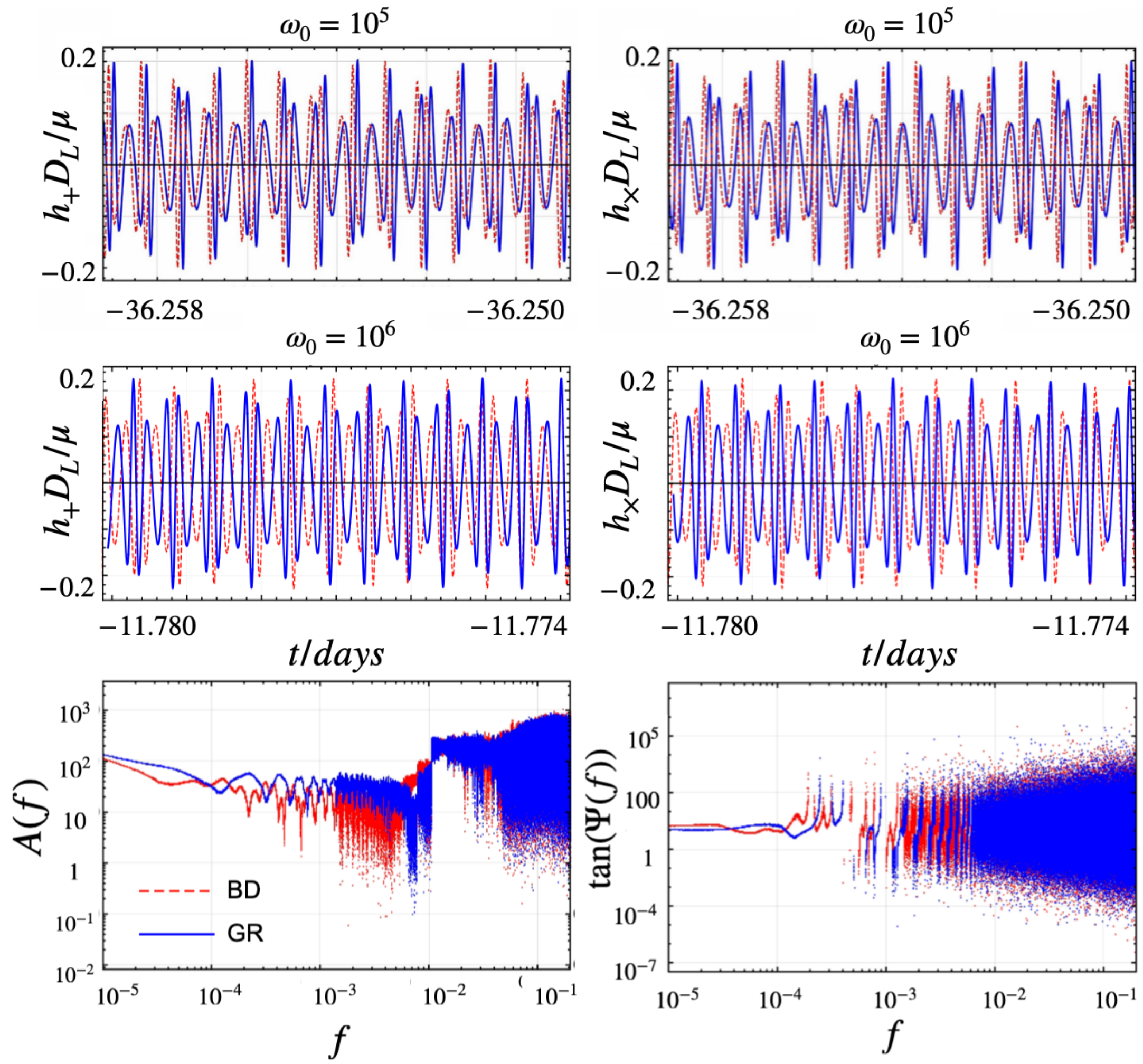}
    \caption{The plus and cross strains $h_+ D_L/\mu$ ($h_\times D_L/\mu$) in the time domain in GR (blue solid lines) and BD theory (red dahsed lines) with different $\omega_0$, $\mu=m_1m_2/(m_1+m_2)$ and $D_L$ are the reduced mass of the binary and the luminosity distance between the GW source and the detector, respectively. Different time windows are chosen for different $\omega_0$ to display the mismatch of the phases in GR and BD theory.
    The zero time is set at the moment that the NS reaches the ISCO.
    In the bottom panel, we show the amplitude $A(f)$ and the phase $\Psi(f)$ of the frequency-domain waveforms in GR and BD theory, where the parameters are the same as the plots in the second row.
  }
    \label{tensormode}
\end{figure*}

\begin{figure*}[htp]
    \centering
     \includegraphics[width=1\linewidth]{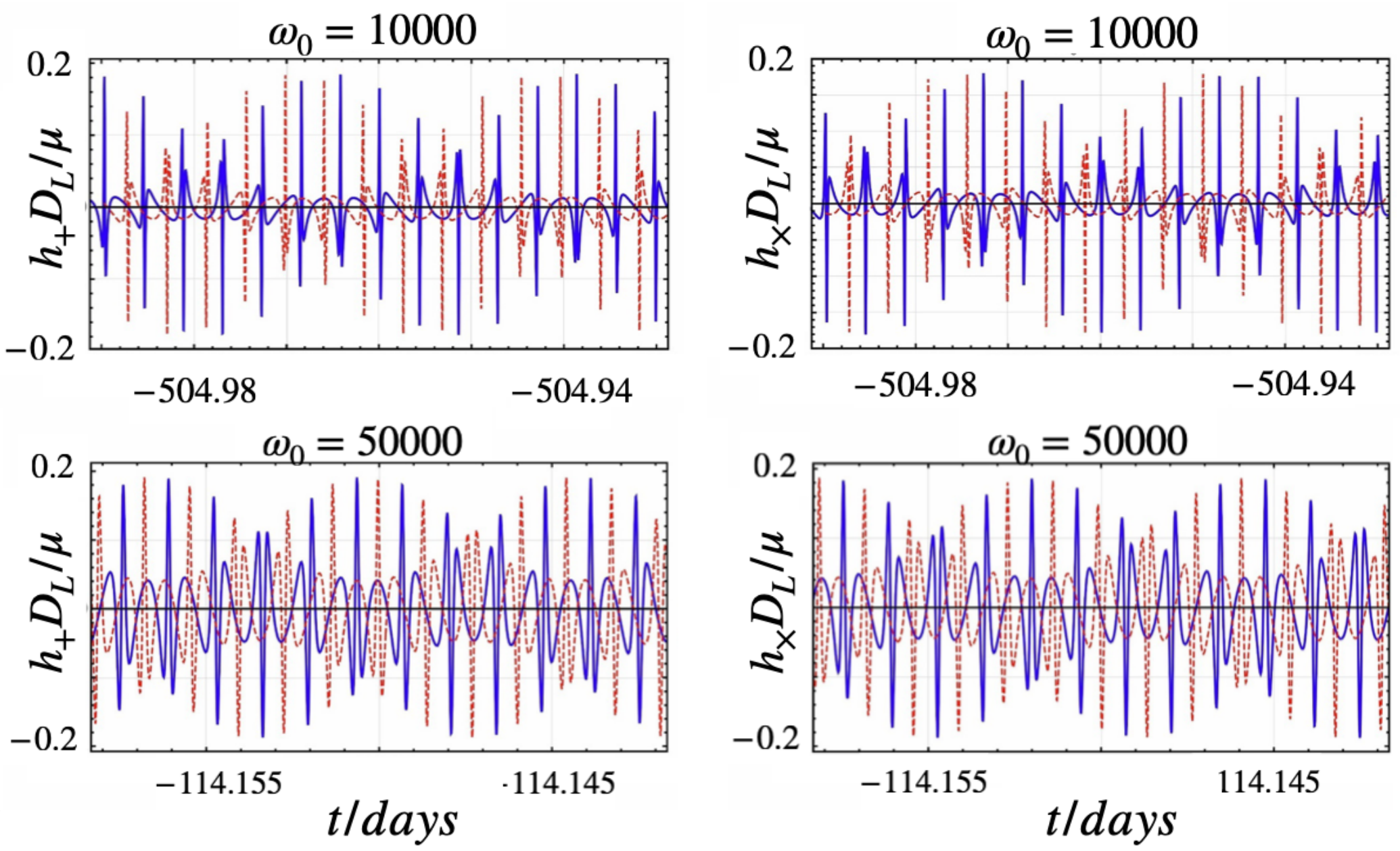}
    \caption{The plus and cross strains $h_+ D_L/\mu$ ($h_\times D_L/\mu$) in the time domain in GR (blue solid lines) and BD theory (red dashed lines) with different $\omega_0$, $\mu=m_1m_2/(m_1+m_2)$ and $D_L$ are the reduced mass of the binary and the luminosity distance between the GW source and the detector, respectively. Different time windows are chosen for different $\omega_0$ to display the mismatch of the phases in GR and BD theory. The zero time is set at the moment that the NS reaches the ISCO.
    }
    \label{tmode}
\end{figure*}

\begin{figure*}[htp]
    \centering
     \includegraphics[width=0.9\linewidth]{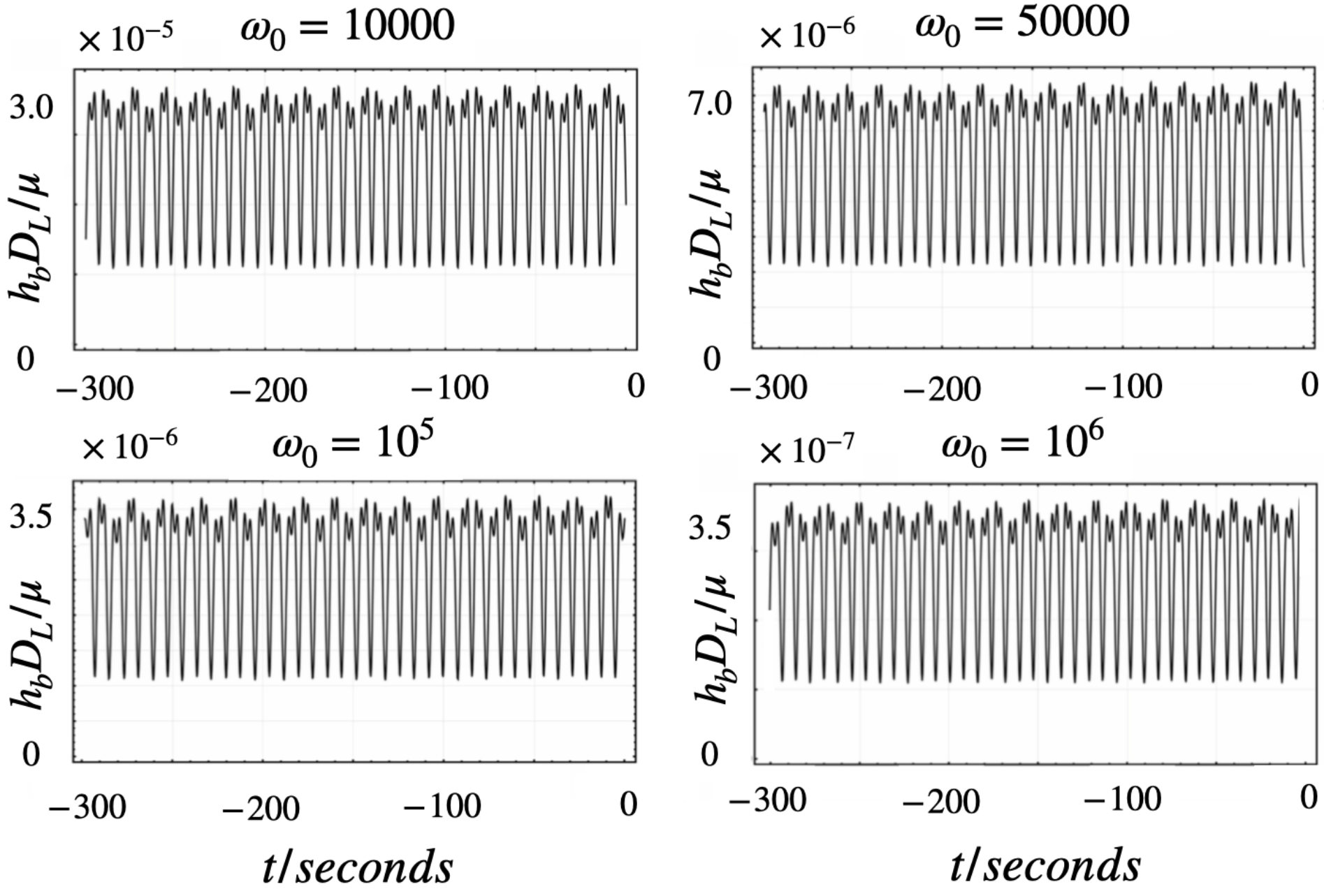}
    \caption{The GW waveform $h_b D_L/\mu$ of the breathing polarization in BD theory with different $\omega_0$, $\mu$ and $D_L$ are the reduced mass of the binary and the luminosity distance between the GW source and the detector, respectively.}
    \label{bmode}
\end{figure*}

We take the inclination angle $\iota=\pi/6$ \cite{Chua:2015mua,Chua:2017ujo}
and consider a two-year inspiral ending at the innermost stable circular orbit (ISCO)
to generate the GW waveforms of EMRIs in GR and BD theory numerically.
Substituting the results of two-year orbital evolution before the ISCO to eqs. \eqref{eq.33}-\eqref{eq.35},
we get the GW waveforms in the time domain
as shown in figures
\ref{tensormode},
\ref{tmode} and \ref{bmode}.
The GW waveforms of plus and cross polarizations in GR and BD theory are shown in figures \ref{tensormode}
and \ref{tmode}.
The GW waveforms of the breathing mode $h_b$ present in BD theory are shown in figure \ref{bmode}.
In figure \ref{tensormode} we show the GW waveforms of the plus and cross polarizations for a binary with $q=10^{-4}$ in GR and BD theory with $\omega_0=10^5$ and $\omega_0=10^6$ in the heliocentric coordinate system.
For the frequency-domain waveforms $\tilde{h}_+=A(f)e^{i\Psi(f)}(1+\cos^2\iota)/2$
and $\tilde{h}_\times=i A(f)e^{i\Psi(f)}\cos\iota$, we show the results for the amplitude $A(f)$ and the phase $\Psi(f)$.
Figures \ref{tensormode}
and \ref{tmode} show that the phase mismatch starts earlier if $\omega_0$ is smaller, i.e., the accumulated phase difference is bigger for the BD theory with smaller $\omega_0$.
As a result, the accumulation of phase difference leads to distinguishable waveforms,
so it is possible to distinguish GR from BD theory with the observations of GWs from I/EMRIs.
From figure \ref{bmode}, we see that the amplitude of the breathing polarization is several orders of magnitude smaller than that of the two tensor polarizations, making it hard to be directly detected.
To detect GWs from I/EMRIs with LISA, the SNR $\rho$ should exceed the threshold value 7 \cite{Cutler:1997ta}.

\section{Detector response}
For the binary system with non-negligible orbital eccentricity, it radiates GWs in multiple harmonics in the inspiral stage. 
In this case, the GWs contain multi-frequency contribution at any moment,
which is very different from the GWs emitted by circular binary
whose frequency is twice the orbital frequency.
When the orbital eccentricity is high,
the high-frequency harmonics become dominant,
the detector response to such multi-band GWs is complicated.

Consider a photon emitted at spacetime event 0, traveling in the direction $\hat{u}_1$,
here a overhat means that it is a unit vector.
It arrives at the end-mirror at spacetime event 1 and then returns at spacetime event 2.
The frequency shift induced by GWs propagating along $\bm{\Omega}$ for this single round trip is \cite{Schilling:1997id,Larson:1999we}
\begin{equation}
\label{deltanu1}
\frac{\Delta\nu(t,\hat{u}_1)}{\nu_0}=\frac{1}{2}\hat{u}_1^i\hat{u}_1^j\left( \frac{h_{ij}^2-h_{ij}^1}{1+\bm{\Omega}\cdot\hat{u}_1}+ \frac{h_{ij}^1-h_{ij}^0}{1-\bm{\Omega}\cdot\hat{u}_1}\right).
\end{equation}
where $h_{ij}^{k}$ is the metric perturbation at spacetime event $k$ ($k=0,1,2$),
\begin{equation}
\label{deltanu2}
\begin{split}
h_{ij}^0&=h_{ij}(t-2L/c),\\
h_{ij}^1&=h_{ij}[t-(L/c)(1+\bm{\Omega}\cdot \hat{u})],\\
h_{ij}^2&=h_{ij}(t),
\end{split}
\end{equation}
$c$ is the speed of light and $L$ is the arm length of the detector.
The phase shift induced by the GWs is
\begin{equation}
\label{laserphase}
\Delta\Phi(t,\hat{u}_1)=2\pi\int_0^t  \Delta\nu(t',\hat{u}_1) dt'.
\end{equation}
The strain recorded in the interferometric detector is 
\begin{equation}
\label{strainH}
H(t)=\frac{c}{4\pi\nu_0L}(\Delta\Phi(t,\hat{u}_1)-\Delta\Phi(t,\hat{u}_2)),
\end{equation}
where $\hat{u}_1$ and $\hat{u}_2$ are the unit vectors along the two arms of the detector.

In this paper, we take LISA as an example to calculate the detector response.
The result can be easily extended to other space-based GW detectors like Tianqin or Taiji.
In the heliocentric coordinate system, the unit vectors of two detector arms, i.e. $\hat{u}_1$ and $\hat{u}_2$ of LISA are \cite{Cutler:1997ta,Zhang:2020drf}:
\begin{equation}
\label{detectorLISA}
\begin{split}
\hat{u}_{1x}&=-\sin(\omega_st ) \cos(\omega_st)  + \cos(\omega_st) \sin(\omega_st)/2,             \\
\hat{u}_{1y}&=     \cos(\omega_st)\cos(\omega_st)+\sin(\omega_st)  \sin(\omega_st)/2,        \\
\hat{u}_{1z}&=\sin(\pi/3) \sin(\omega_st),         \\
\hat{u}_{2x}&=-\sin(\omega_st ) \cos(\omega_st-\pi/3)+ \cos(\omega_st) \sin(\omega_st-\pi/3)/2,             \\
\hat{u}_{2y}&=     \cos(\omega_st)\cos(\omega_st-\pi/3)+ \sin(\omega_st)  \sin(\omega_st-\pi/3)/2,        \\
\hat{u}_{2z}&=\sin(\pi/3) \sin(\omega_st-\pi/3),         \\
\end{split}
\end{equation}
where the rotation frequency $\omega_s=2\pi/(365 $ days) and we set the initial phase to be zero.
When we calculate the response of the detector in the heliocentric coordinate,
we have to consider the phase modulation induced by the translatory motion of the detector 
which introduces an extra time delay
\begin{equation}
    t_d(t)=\frac{  \bm{\Omega} \cdot \hat{d}_{\text{LISA}}(t) }{c},
\end{equation}
where we adopt $\hat{d}_{\text{LISA}}(t)=(\cos\omega_s t,\sin \omega_s t,0)\times1 \text{AU}$ for simplicity.
Then the frequency shift in one arm becomes
\begin{equation}
\label{deltanu3}
\begin{split}
\frac{\Delta\nu(t,\hat{u}_1)}{\nu_0}=\frac{1}{2}\hat{u}^i_1\hat{u}^j_1&\left( \frac{h_{ij}(t-t_d)-h_{ij}(t-t_d-(L/c)(1+\bm{\Omega}\cdot\hat{u}_1))}{1+\bm{\Omega}\cdot\hat{u}_1} \right. \\
&\qquad\qquad+\left. \frac{h_{ij}(t-t_d-(L/c)(1+\bm{\Omega}\cdot\hat{u}_1))-h_{ij}(t-t_d-2L/c)}{1-\bm{\Omega}\cdot\hat{u}_1}\right).
\end{split}
\end{equation}
The frequency shift in the other arm can be obtained by replacing $\hat{u}_1$ with $\hat{u}_2$.

Substitute eqs.~\eqref{hijgr} and \eqref{hijbd} into eqs. \eqref{laserphase}, \eqref{strainH} and \eqref{deltanu3}, we obtain the strain $H_{\text{GR}}(t)$ for GR and $H_{\text{BD}}(t)$ for BD theory, respectively.

With a signal $H(t)$, the SNR in LISA is
\begin{equation}
\label{snr}
\begin{split}
\rho ^2&=4\int_{f_1}^{f_2}df\frac{1}{S_{n}(f)} \tilde{H}(f)\tilde{H}^*(f),
\end{split}
\end{equation}
where $\tilde{H}(f)$ is the Fourier transform of the signal $H(t)$,
$f_2$ is the frequency at the ISCO and $f_1$ is the frequency two years before the ISCO,
the noise power spectral density $S_n(f)$ of LISA is \cite{Cornish:2018dyw}
\begin{equation}
\label{noisepsd}
S_n(f)=\frac{S_x}{L^2}+\frac{2S_a\left(1+\cos^2(2\pi f L/c) \right)}{(2\pi f)^4L^2}
\left(1+\left(\frac{4\times10^{-4}\text{Hz}}{f}\right)^2\right),
\end{equation}
the acceleration noise is $\sqrt{S_a}=3\times 10^{-15}\,\text{m s}^{-2}/\text{Hz}^{1/2}$,
the displacement noise is $\sqrt{S_x}=1.5\times 10^{-11}\,\text{m/Hz}^{1/2}$ and the arm length is $L=2.5 \times 10^6\,\text{km}$ \cite{Audley:2017drz}.

Take the luminosity distance of the EMRI as $D_L=100$ Mpc \cite{Gondan:2018khr,Ma:2019rei}  and use the GW waveforms obtained in figures
\ref{tensormode},
\ref{tmode} and \ref{bmode},
we calculate the signals $H_\text{GR}(t)$ and $H_\text{BD}(t)$ registered in LISA for GR and BD theory,
then we calculate the Fourier transforms $\tilde{H}_{\text{GR}}(f)$ and $\tilde{H}_{\text{BD}}(f)$ of $H_\text{GR}(t)$ and $H_\text{BD}(t)$, respectively.
With a two-year observation of a NS/BH binary with $q=10^{-4}$ ($q=10^{-3}$),
we get the SNR $\rho(\tilde{H}_{\text{GR}})\approx 23.89$ (13.22) for GR and $\rho(\tilde{H}_{\text{BD}})\approx 23.92$ (13.22) for the BD theory with $\omega_0=10^6$.
Thus, these GW signals are detectable for LISA-like space-based detectors.
To quantify the difference of GWs between different theories,
following \cite{Kocsis:2011dr},
we compute the SNR $\rho$ of the difference between these two signals $\Delta \tilde{H}=\tilde{H}_\text{BD}-\tilde{H}_\text{GR}$ and we get $\rho(\Delta \tilde{H})\approx 33.86 \ (17.97)>10$, which means the difference between the waveforms is significant.
Therefore, LISA can detect GWs from a binary consisting of a NS with the mass $m_2=1.3M_\odot$ and a BH with the mass $m_1=10^4 m_2 \ (10^3 m_2)$ located at $D_L=100$ Mpc away, and a two-year observation of the binary can place the bound $\omega_0>10^6$.

On the other hand, the overlap of two GW waveforms can be quantified by \cite{Babak:2006uv}
\begin{equation}
    \mathcal{O}(\tilde{H}_1,\tilde{H}_2)= \frac{(\tilde{H}_1|\tilde{H}_2)}{\sqrt{(\tilde{H}_1|\tilde{H}_1)(\tilde{H}_2|\tilde{H}_2)}},
\end{equation}
where the inner product$(a(f)|b(f))$ is defined as
\begin{equation}
    (a(f)|b(f))=2\int^{\infty}_{0} df \frac{\tilde{a}^*(f) \tilde{b}(f)+\tilde{a}(f)\tilde{b}^*(f)}{S_n(f)}.
\end{equation}
For two identical GW waveforms, the overlap is exactly 1.
The mismatch between two waveforms is \cite{Apostolatos:1995pj,Allen:2005fk,LIGOScientific:2016ebw,Datta:2019epe,Bozzola:2020mjx}
\begin{equation}
    \text{mismatch}(\tilde{H}_1,\tilde{H}_2)=1- \max \mathcal{O}(\tilde{H}_1,\tilde{H}_2),
\end{equation}
where the maximum is evaluated with respect to time-shift and orbital-phase shift.
Two waveforms are considered experimentally indistinguishable if their mismatch is smaller than $d/(2\rho^2)$ wtih $d$ being the number of parameters \cite{Flanagan:1997kp,Lindblom:2008cm,McWilliams:2010eq,Chatziioannou:2017tdw}.
Take $q=10^{-4}$, $d=10$ and $\omega_0=10^6$, we find the mismatch of the waveforms between GR and BD is 0.5, which is much larger than $d/(2\rho^2(\tilde{H}_{\text{GR}}))=8.8\times10^{-3}$.
Thus these two waveforms can be distinguished from each other.
The results are consistent with the analysis above by calculating the SNR of the difference between the signals.
We summarize the results in table \ref{t-s1}.

\section{Conclusion}
As the mass of the central black hole increases,
the NS starts closer to the central black hole and stays in stronger field regions,
so the constraint on BD theory becomes weaker.
Compared with circular orbits, the NS starts closer in eccentric orbits,
so eccentric orbits don't always give stronger constraint on BD theory.
For $q=10^{-3}$, eccentric binaries can give stronger constraint,
but for $q<10^{-4}$, the constraint by eccentric binaries is weaker.
Even considering the degeneracy among 10 parameters, EMRIs with $q=10^{-6}$ can give the constraint $\omega_0>10^6$ because the mismatch 0.26 is much larger than $10/[2\times(152.45)^2]=2.2\times10^{-4}$.
For IMRIs with $q=10^{-3}$, we get $\rho(\tilde{H}_{\text{GR}})\approx 12.49$, $\rho(\tilde{H}_{\text{BD}})\approx 12.49$ with $\omega_0=10^6$ and $\rho(\Delta \tilde{H}) \approx 17.97$,
the mismatch is 0.73 which is larger than $d/(2\rho^2)=0.03$ with $d=10$,
so we get the bound $\omega_0>10^6$.
We also consider a two-year observation of EMRIs with $q=10^{-4}$ by LISA starting from $p_0=80$
and find that the results are similar.

\begin{table}
\begin{tabular}{ |p{3.5cm}|| c c c c | }
    \hline
    System  & $\rho(\tilde{H}_\text{GR})$  &$\rho(\tilde{H}_\text{BD})$  &$\rho(\Delta \tilde{H})$  & Mismatch \\
    \hline
    $q=10^{-3}$   &    &    &    &\\
    Circular  & 12.49   & 12.49\ (12.49)  & 17.88\ (17.65)   & 0.98\ (0.71) \\
    Eccentric ($e_0=0.8$)   & 13.22   & 13.22\ (13.22)   & 18.69\ (17.97)   & 0.99\ (0.73)  \\
    \hline
    $q=10^{-4}$  &   &   &   & \\
    Circular  & 26.86   & 26.86\ (26.87)   & 38.41\ (34.48)    & 0.97\ (0.56)  \\
    Eccentric ($e_0=0.8$) & 23.89   & 23.88\ (23.92)    & 34.16\ (33.86)   & 0.97\ (0.5) \\
    \hline
    $q=10^{-6}$  &   &   &   & \\
    Circular  & 197.94   & 197.95\ (197.94)   & 284.12\ (278.51)    & 0.94\ (0.33)  \\
    Eccentric ($e_0=0.8$) & 152.45   & 152.45\ (152.44)    & 218.14\ (211.11)   & 0.90\ (0.26) \\
    \hline
\end{tabular}
\caption{The results of SNR and mismatch for circular and eccentric orbits in GR and BD theory with a two-year integration prior to the ISCO. We take $\omega_0=10^5$ and $\omega_0=10^6$, and the results in parentheses correspond to $\omega_0=10^6$.}
\label{t-s1}
\end{table}

Because the waveform templates we constructed are more accurate, our results should be more robust
than the crude estimations derived with the analytic PN waveform and the FIM approximation \cite{Scharre:2001hn,Will:2004xi,Berti:2004bd,Yagi:2009zm}.
Our results show that with a two-year observation of IMRIs/EMRIs,
we can get the constraint $\omega_0>10^6$ which is stronger than the current constraint obtained by the solar system tests  \cite{Bertotti:2003rm}.
For real observational data with Bayesian analysis, the constraint may become weaker due to the degeneracies among parameters.

\begin{acknowledgments}
TJ and YG are grateful to Adam Pound for helpful discussion on the SF method.
The numerical computations were performed at the public computing service platform provided by Network and Computing Center of HUST.
This research is supported in part by the National Key Research and Development Program of China under Grant No. 2020YFC2201504,
the NSFC under Grant Nos. 11875136 and 12147120,
and China Postdoctoral Science Foundation under Grant No. 2021TQ0018.
\end{acknowledgments}


\providecommand{\href}[2]{#2}\begingroup\raggedright\endgroup

\end{document}